\documentclass[twocolumn,showpacs,preprintnumbers,amsmath,amssymb]{revtex4}
\usepackage{graphicx}
\usepackage{dcolumn}
\usepackage{bm}

\newcommand{\be}{\begin{equation}}
\newcommand{\ee}{\end{equation}}
\newcommand{\ba}{\begin{eqnarray}}
\newcommand{\ea}{\end{eqnarray}}

\begin{document}
\preprint{}
\title{The $\gamma \gamma$ decay of the $f_0(1370)$ and $f_2(1270)$ resonances 
in the hidden gauge formalism.}

\author{
H. Nagahiro$^1$\footnote{Present address: Department of Physics, Nara Women's University, Nara 630-8506, Japan}, J. Yamagata-Sekihara$^2$, E. Oset$^{1,3}$, S. Hirenzaki$^4$ and R. Molina$^3$
}

\affiliation{$^1$Research Center for Nuclear Physics (RCNP), Osaka University, Ibaraki, Osaka 567-0047, Japan \\
$^2$Yukawa Institute for Theoretical Physics, Kyoto University, Kyoto 606-8502, Japan\\
$^3$ Departamento de F\'{\i}sica Te\'orica and IFIC,
Centro Mixto Universidad de Valencia-CSIC,
Institutos de
Investigaci\'on de Paterna, Apartado 22085, 46071 Valencia, Spain\\
$^4$ Department of Physics, Nara Women's University, Nara 630-8506, Japan}%

\date{\today}


 \begin{abstract} 
 
 Using recent results obtained within the hidden gauge formalism for vector
 mesons, in which the $f_0(1370)$ and $f_2(1270)$ resonances are dynamically
 generated resonances from the $\rho \rho$ interaction, we evaluate the
 radiative decay of these resonances into $\gamma \gamma$. We obtain results for the width in
 good agreement with the experimental data for the $f_2(1270)$ state and 
 a width about a factor two smaller for the $f_0(1370)$ resonance, which is also in agreement with the data of the Crystal Ball collaboration and with the more recent ones from the Belle collaboration, which however have a very large uncertainty.

\end{abstract}


\maketitle

\section{Introduction}

The radiative decay of mesons has been traditionally advocated as a good tool
to investigate the nature of the controversial mesons. In particular the decay
into $\gamma \gamma$ has had a special attention \cite{mike}. The scalar mesons
have been those most thoroughly studied given the on going debate on whether
they are $q \bar{q}$ states, tetraquarks, or meson-meson molecules as a
particular case of the more general one, dynamically generated states from the
meson-meson interaction in coupled channels, see \cite{mikerecent} for a recent
review.  

  In this work we want to call the attention to two particular mesons, the 
$f_0(1370)$ and $f_2(1270)$ states, because in a recent paper \cite{raquel} the
two mesons were found as dynamically generated states from the $\rho-\rho$ 
interaction
in the hidden gauge approach for the vector mesons \cite{hidden1,hidden2}. The
attraction was stronger in the spin S=2 channel than in the scalar one, but in
both channels there was enough attraction to generate bound states. In other
channels the interaction was either very weak or repulsive, such that these two
states stand as particular cases which can be viewed as largely being $\rho-\rho$
molecules in that framework. It is interesting to see how this idea immediately leads one to
evaluate rather accurately the partial decay width of the two states into 
$\gamma \gamma$ and this is the purpose of this paper.

Although Ref. \cite{raquel} contains the first theoretical evaluation of the
$\rho \rho$ bound system, it is interesting to recall that, based on
phenomenological properties (the large $\Gamma _{\rho \rho}$ versus  $\Gamma
_{\eta \eta}$), the $\rho \rho$ molecular nature of the $f_0(1370)$ was also
suggested in  \cite{klempt,crede}. The $f_2(1270)$ is, however, widely believed
to be part of a p-wave nonet of $q \bar{q}$ states \cite{klempt,crede}. The
results of \cite{raquel} support the suggestion of \cite{klempt,crede} for the
$f_0(1370)$ as a $\rho \rho$ molecule, but surprisingly also show that for
spin S=2 the $\rho \rho$ interaction is attractive and about three times
larger than in the case of S=0, thanks to which a stronger bound $\rho \rho$ 
state appears for S=2, which was identified in \cite{raquel} as the $f_2(1270)$
resonance.

  The experimental situation is rich in the case of the $f_2(1270)$, which has
a very pronounced peak in $\gamma \gamma$ scattering going to pions.
Compatible results are found in  different laboratories and using 
different methods,
Crystal Ball \cite{Marsiske:1990hx}, Mark II \cite{Boyer:1990vu}, 
JADE \cite{Oest:1990ki}, TOPAZ \cite{Adachi:1989dd}, MD-1 \cite{Blinov:1992pr},
CELLO \cite{Behrend:1992hy} and VENUS \cite{Yabuki:1995ai}. The PDG quotes the
result $\Gamma(f_2(1270) \to \gamma \gamma)= 2.71 ^{+ 0.26} _{-0.23}$ keV
\cite{Amsler:2008zz}.
 Recent results are also presented by the Belle collaboration 
in \cite{Mori:2007bu} and in \cite{Pennington:2008xd}, where the preferred
solution gives $\Gamma(f_2(1270) \to \gamma \gamma)= 3.14\pm 0.20$ keV.
The situation of the $\gamma \gamma $ decay of the $f_0(1370)$ is rather
unclear. The latest edition of the PDG \cite{Amsler:2008zz} does not quote any
value, superseding old results which  were  ambiguous.
The Belle collaboration has the most recent results in this direction~\cite{uehara}.
This work quotes a central value for the mass of the $f_0(1370)$ of 1470 MeV, but with very large uncertainties, of the order of 255 MeV, mostly of systematic origin.
It also quotes a value for the radiative decay of this resonance, again with a very large uncertainty, $\Gamma_{\gamma \gamma} B(\pi^0 \pi^0)$=($11^{+4+603}_{-2-7}$) eV.
The same work quotes more accurate values deduced by the Crystal Ball collaboration~\cite{Marsiske:1990hx}, $\Gamma_{\gamma \gamma} B(\pi^0 \pi^0)$=($430\pm 80$) eV, with, however, much less statistics than Belle.

 On the theoretical side, an evaluation of the radiative decay into $\gamma
 \gamma$ of the $f_0(1370)$  has been done in \cite{Rodriguez:2004tn}, where
 using a model where the scalars are a mixture of $q \bar{q}$ and 
 $qq \bar{q} \bar{q}$ the authors find a small value between $0-0.22$ keV.
 Much bigger values, of the order of $4$~keV are obtained in \cite{eef} 
 assuming the state to be basically a $q \bar{q}$ of non strange nature,
 although actually the value quoted is used as input to determine parameters
 of the theory.
 Ratios of radiative widths between scalar states are also quoted in 
 \cite{Close:1996sv} under the assumption that they are $q \bar{q}$ states 
 mixing with glueballs . Results for the $\gamma \gamma$ decay of the 
 $f_2(1270)$ state are also obtained in \cite{Anisovich:2001zp}, where assuming 
 that the resonance 
 is a  $q \bar{q}$ state, a satisfactory description of this decay rate together
 with that of the $f_2(1525)$ is obtained at the expense of fitting two free
 parameters.

   The novel picture of \cite{raquel} puts the $f_0(1370)$ and $f_2(1270)$
resonances on the same footing, allowing one to calculate the $\gamma \gamma$
radiative width within the same formalism. This is the aim of the present paper.
The evaluation presented here turns out to be rather simple technically, once the
formalism for the generation of the two resonances is developed in
\cite{raquel}. We will find that the widths obtained are rather precise, with respect to uncertainties from the parameters of the model, and agree well with the well known experimental results for the case of the 
$f_2(1270)$, while  the one for the  
$f_0(1370)$ follows the actual experimental trend that it is indeed about one 
order of magnitude smaller than that for the $f_2(1270)$ state. 

\section{Formalism}
\begin{figure}[htpd]
\begin{center}
\includegraphics[width=7cm]{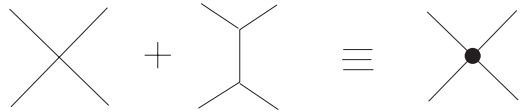}
\caption{\label{fig:1}
Driving terms of the $\rho\rho$ interaction.
The diagrams to the right sums the contribution of the first two diagrams.} 
\end{center}
\end{figure}

\begin{figure}[htpd]
\begin{center}
\includegraphics[width=7.5cm]{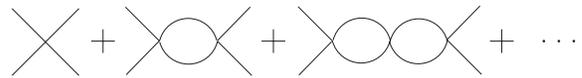}
\caption{\label{fig:2}
Diagrams summed up in the Bethe Salpeter equation.} 
\end{center}
\end{figure}

\begin{figure*}[htpb]
\begin{center}
\includegraphics[width=17cm]{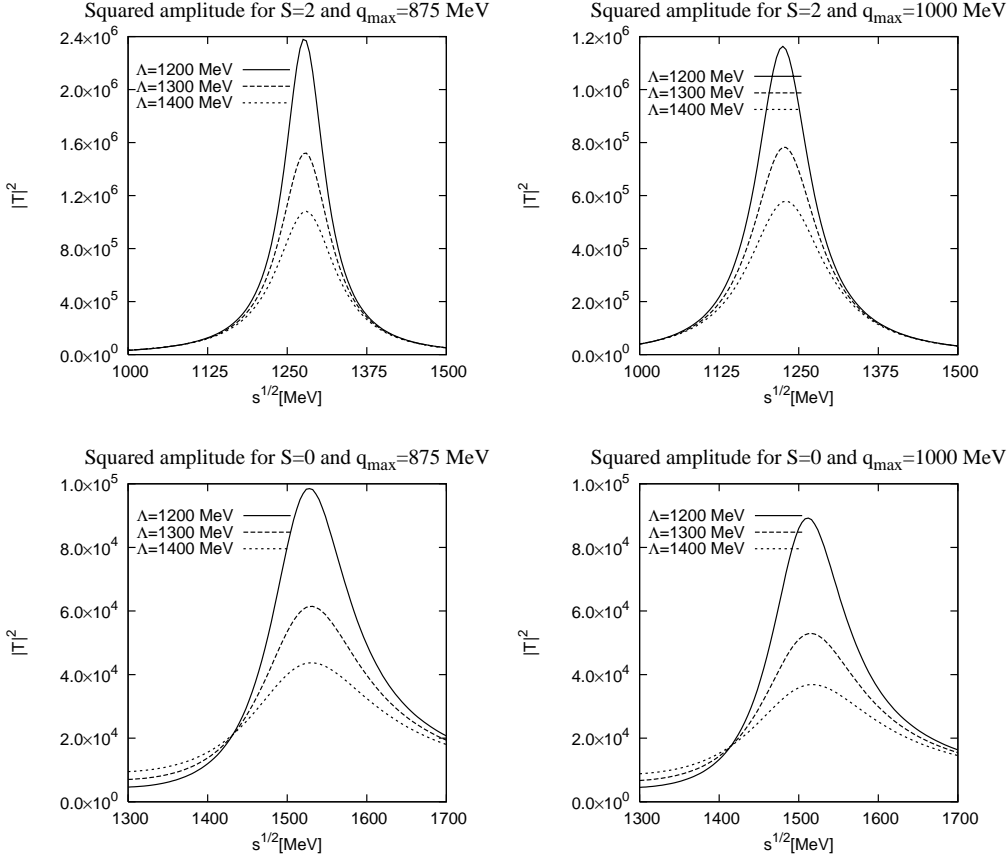}
\caption{\label{fig:3}
$|T|^2$ calculated in~\cite{raquel} for $S=0$ and $S=2$ for several values of $\Lambda$ and $q_{\rm max}$ defined in \cite{raquel}.
}
\end{center}
\end{figure*}

In \cite{raquel} the driving term for the $\rho\rho$ interaction was obtained from the hidden gauge Lagrangian
\begin{equation}
\label{eq:1}
{\cal L}_{III}=-\displaystyle{\frac{1}{4}}\langle V_{\mu \nu}V^{\mu \nu}\rangle~~,
\end{equation}
where the symbol $\langle$ $\rangle$ stands for the $SU(3)$ trace and $V_{\mu\nu}$ is given by
\begin{equation}
\label{eq:2}
V_{\mu\nu}=\partial_\mu V_\nu-\partial_\nu V_\mu -ig[V_\mu,V_\nu]~~,
\end{equation}
with $g=M_V/2f$, and $f=93$ MeV is the pion decay constant.
The $SU(3)$ matrix of $V_\mu$ is given by
\begin{eqnarray}
\label{eq:3}
V_\mu=
\bordermatrix{
&~&~\cr
&\frac{\rho^0}{\sqrt 2}+\frac{\omega}{\sqrt 2}&\rho^+&K^{*+}\cr
&\rho^-&-\frac{\rho^0}{\sqrt 2}+\frac{\omega}{\sqrt 2}&K^{*0}\cr
&K^{*-}&{\bar K^{*0}}&\phi\cr
}
_\mu~~.
\end{eqnarray}
The interaction of ${\cal L}_{III}$ of eq.~(\ref{eq:3}) gives rise to a contact term
\begin{equation}
\label{eq:4}
{\cal L}_{III}^{(c)}=\displaystyle{\frac{g^2}{2}}\langle V_\mu V_\nu V^\mu V^\nu-V_\nu V_\mu V^\mu V^\nu \rangle~~
\end{equation}
and a three vector vertex given by
\begin{equation}
\label{eq:5}
{\cal L}_{III}^{(3V)}=ig\langle (\partial_\mu V_\nu - \partial_\nu V_\mu)V^\mu V^\nu \rangle~~.
\end{equation}
With this information the driving term for the $\rho\rho$ interaction is given by the diagrams of fig.~\ref{fig:1}.
This driving term, $V$, is used as kernel in the Bethe Salpeter equation depicted in fig.~\ref{fig:2},
which gives the solution
\begin{equation}
\label{eq:6}
T=\displaystyle{\frac{V}{1-VG}}~~,
\end{equation}
with $G$ the loop function for the meson propagators conveniently regularized \cite{raquel}.
The interaction is studied for $I=0$ and the projections over spin and isospin are performed.
Two states are obtained, visible in neat peaks of $|T|^2$, which is depicted in fig.~\ref{fig:3}.
They correspond to the $f_2(1270)$ and $f_0(1370)$, the later one appearing around $1500$ MeV in our approach, close to the preliminary results of the Belle collaboration \cite{uehara}.

The model of \cite{raquel} contains $\rho\rho$ as basic components to form the scalar and tensor states.
However, intermediate $\pi\pi$ states, through the box and crossed box diagrams, were also considered.
In addition, intermediate $\omega\omega$ states, driven by pion exchange through anomalous $\rho\omega\pi$ couplings, were also taken into account.
It was found there that the real parts of the $\pi\pi$ and $\omega\omega$ intermediate states mechanisms were individually small compared to the dominant tree level $\rho\rho\to \rho\rho$ mechanisms and in addition there were cancellations between the $\pi\pi$ and $\omega\omega$ contributions, rendering the tree level $\rho\rho$ terms largely dominant.

The calculations of Ref. \cite{raquel} were done using the on-shell approach of 
\cite{ollerulf} based on the N/D method, using a cut off in the three momentum in
the loops, which was shown in \cite{ollerulf} to be equivalent to the use of
dimensional regularization. This prescription then preserves the underlying
symmetries and gauge invariance ( see a more detailed discussion in
\cite{mishadan}, page 5). The approach of Ref. \cite{raquel} uses a full
relativistic treatment of the loop functions, which guarantees exact unitarity
and analiticity of the amplitudes. Nonrelativistic approximations are done in
the evaluation of the $VV$ potential, neglecting the three momentum of the
vector mesons versus their mass. While this approximation is  quite good for
the $f_0(1370)$ state, for the case of the more bound $f_2(1270)$ resonance
certainly it induces a larger correction, still under control as discussed in
Ref. \cite{raquel} ( see pag 4 of this reference), particularly because a small
fine tuning of the parameters is allowed in the approach to fit one resonance
mass, which allows one to cope with small correcions stemming from different
sources.

Figure~\ref{fig:3} shows results including also the box diagram accounting for $\pi\pi$ decay, which plays a minor role in the binding of the two states but enlarges the width of the states due to the large phase space available for decay into two pions.
The $\Lambda$ parameter in fig.~\ref{fig:3} appears to account for $\rho \rightarrow \pi\pi$ off shell and is varied within reasonable values~\cite{Titov:2000bn}.

\begin{figure}[htpd]
\begin{center}
\includegraphics[width=7.5cm]{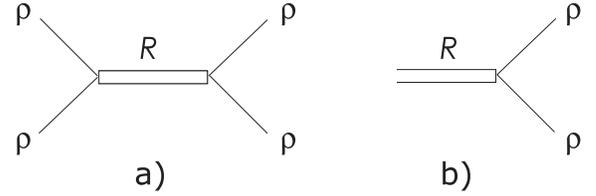}
\caption{\label{fig:4}
a) Resonance pole representation of the amplitude of \cite{raquel}.
b) Diagram depicting the coupling of the resonance to $\rho\rho$.}
\end{center}
\end{figure}

The amplitude of fig.~\ref{fig:2} can be parameterized as a Breit-Wigner amplitude, and using the spin projection operators of \cite{raquel} we find 
\begin{eqnarray}
S=2&&\nonumber\\
t^{(2)}&=&\displaystyle{\frac{g_T^2}{s-M_R^2+iM_R\Gamma}}\nonumber\\
&\times&\{\displaystyle{\frac{1}{2}(\epsilon^{(1)}_i\epsilon^{(2)}_j+\epsilon^{(1)}_j\epsilon^{(2)}_i)-\frac{1}{3}\epsilon^{(1)}_l\epsilon^{(2)}_l\delta_{ij}}\}\nonumber\\
&&\{\displaystyle{\frac{1}{2}(\epsilon^{(3)}_i\epsilon^{(4)}_j+\epsilon^{(3)}_j\epsilon^{(4)}_i)-\frac{1}{3}\epsilon^{(3)}_m\epsilon^{(4)}_m\delta_{ij}}\}
\label{eq:7}\\
S=0&&\nonumber\\
t^{(0)}&=&\displaystyle{\frac{g_S^2}{s-M_R^2+iM_R\Gamma}}\times
\displaystyle{\frac{1}{\sqrt {3}}\epsilon^{(1)}_i\epsilon^{(2)}_i\frac{1}{\sqrt 3}\epsilon^{(3)}_j\epsilon^{(4)}_j}
\label{eq:8}
\end{eqnarray}
where $\epsilon^{(m)}_i$ are the polarization vectors of the $\rho$ for each $m$ of the four $\rho$ mesons involved (1, 2 for the initial states and 3, 4 for the final states).
As shown in \cite{raquel}, because of the small three momenta of the $\rho$ mesons involved, only the spatial components of the $\rho$ polarization vectors are needed.
These amplitudes correspond to a pole term as depicted in fig.~\ref{fig:4}.
Since we are interested in the coupling of the resonance to the $\rho\rho$ system, this is given by
\begin{eqnarray}
S=2&&\nonumber\\
&&g_T[\displaystyle{\frac{1}{2}(\epsilon^{(3)}_i\epsilon^{(4)}_j+\epsilon^{(3)}_j\epsilon^{(4)}_i)-\frac{1}{3}\epsilon^{(3)}_m\epsilon^{(4)}_m\delta_{ij}}]
\label{eq:9}\\
S=0&&\nonumber\\
&&g_S\frac{1}{\sqrt 3}\epsilon^{(3)}_i\epsilon^{(4)}_i~~.
\label{eq:10}
\end{eqnarray}
In both cases we are only interested in the isospin $I=0$ component, given by
\begin{equation}
\label{eq:11}
|\rho\rho~I=0\rangle=-\frac{1}{\sqrt 6}|\rho^+\rho^-+\rho^-\rho^++\rho^0\rho^0\rangle~~,
\end{equation}
which uses the unitary normalization (extra $\frac{1}{\sqrt 2}$ factor to account for identical particles in the sum over intermediate states) and the phase convention $|\rho^+\rangle=-|1,1\rangle$.
The use of this normalization will also account for the factor $1/2$ of symmetry that one has when dealing with identical particle in the final state.

We only need the $\rho^0\rho^0$ component of the amplitude $R\rightarrow \rho\rho$.
The $\rho^0\rho^0$ component is given by $(-1/{\sqrt 3})$ times the $I=0$ components of eqs.~(\ref{eq:9}), (\ref{eq:10}).

\section{Radiative decays}
Here we present the formalism for the $\gamma\gamma$ decay of the two resonances.
Since the resonances are formed from $\rho\rho$ components, the two photons are radiated from these components.
This is taken into account by loops involving the $\rho$ mesons, in a similar way as done in \cite{hideko} for the radiative decay into $\pi\gamma$ of the axial vector mesons generated dynamically from the interaction of vectors and pseudoscalars within the same hidden gauge formalism.

\begin{figure}[htpd]
\begin{center}
\includegraphics[width=7.5cm]{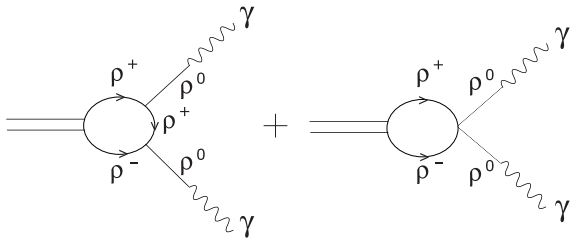}
\caption{\label{fig:5}
Feynman diagrams to evaluate the radiative decay width of $f_0(1370)$ and $f_2(1270)$.}
\end{center}
\end{figure}

\begin{figure}[htpd]
\begin{center}
\includegraphics[width=7.5cm]{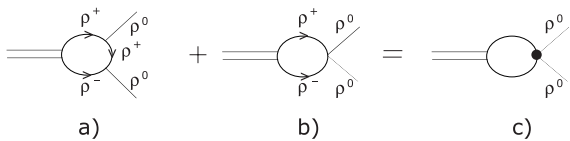}
\caption{\label{fig:6}
Strong part of feynman diagrams to evaluate the radiative decay width of $f_0(1370)$ and $f_2(1270)$.}
\end{center}
\end{figure}

Taking into account that in the hidden gauge formalism the photons do not couple directly to the vector but indirectly through their conversion into $\rho$, $\omega$, $\phi$, the picture we want for the $\gamma\gamma$ decay of the resonances is given in fig.~\ref{fig:5}.
The fact that the photon couples to vectors by direct conversion into another vector, allows one to factorize the diagrams of fig.~\ref{fig:5} into a strong part, $R\rightarrow \rho^0\rho^0$, depicted in fig.~\ref{fig:6}, followed by the photon coupling to either $\rho^0$.
Note that whether we have the strong interaction terms, or the electromagnetic ones of fig.~\ref{fig:5}, the loop contains $\rho^+\rho^-$ alone since both, the $\rho^0\rho^0\rho^0\rho^0$ contact term of the $\rho^0\rho^0\rho^0$ three leg vertex are zero.

Coming back to the diagrams of fig.~\ref{fig:6}, we see that by definition of the potential or kernel of the interaction, see fig.~\ref{fig:1}, the sum of the first two diagrams can be cast as the diagram of fig.~\ref{fig:6}c, which is given by
\begin{equation}
t_{R\rho^0\rho^0}=\displaystyle{(-\frac{1}{\sqrt 3})}g^{(i)}G(M_R)VP^{(i)}~~,
\label{eq:12}
\end{equation}
where $g^{(i)}$ stands for $g^{(S)}$ or $g^{(T)}$, $P^{(i)}$ are the corresponding spin operators of eqs.~(\ref{eq:9}), (\ref{eq:10}) and $G(M_R)$ stands for the loop function defined in eq.~(\ref{eq:6}) evaluated at $\sqrt{s}=M_R$.

\begin{figure}[htpb]
\begin{center}
\includegraphics[width=4.5cm]{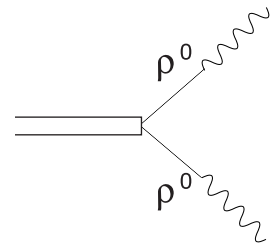}
\caption{\label{fig:7}
Feynman diagram equivalent to those of fig.~\ref{fig:5} at the resonance pole energy.}
\end{center}
\end{figure}

However, according to eq.~(\ref{eq:6}), we are now at the pole of the amplitude, where $GV=1$, and, thus, we obtain
\begin{equation}
\label{eq:13}
t_{R\rho^0\rho^0}=-\displaystyle{\frac{1}{\sqrt 3}}g^{(i)}P^{(i)}
\end{equation}
which is the same coupling as in eqs.~(\ref{eq:9}), (\ref{eq:10}) including the isospin factor for $\rho^0\rho^0$.
In other words, the addition of an extra bubble (loop) to the series of diagrams of fig.~\ref{fig:2} leads to the same series at the pole of the resonance.
This means that the coupling of two photons to the resonance is given by the diagram of fig.~\ref{fig:7}.
Namely, that in the present case, and due to the peculiar couplings of the hidden gauge formalism, the coupling of $\gamma\gamma$ to the dynamically generated $\rho\rho$ resonances is given by the tree level diagram of fig.~\ref{fig:7} alone.
This makes the evaluation obviously very simple, and taking into account the coupling of the photon to the $\rho^0$ \cite{hidden1,hidden2,hideko}
\begin{equation}
\label{eq:14}
-it_{\rho^0\gamma}=(-i)\displaystyle{\frac{1}{\sqrt 2}M_V^2\frac{e}{g}}\epsilon_\mu(\rho)\epsilon^\mu(\gamma)~~~~(e<0)~~,
\end{equation} 
we find at the end the two amplitudes
\begin{eqnarray}
S=2&&\nonumber\\
t_{R\rightarrow \gamma\gamma}&=&-\displaystyle{\frac{1}{\sqrt 3}\frac{e^2}{2}\frac{g_T}{g^2}}\nonumber\\
&\times&\displaystyle{[\frac{1}{2}(\epsilon_i(\gamma_1)\epsilon_j(\gamma_2)+\epsilon_j(\gamma_1)\epsilon_i(\gamma_2))}\nonumber\\
&&\displaystyle{-\frac{1}{3}\epsilon_m(\gamma_1)\epsilon_m(\gamma_2)\delta_{ij}]}
\label{eq:15}\\
S=0&&\nonumber\\
t_{R\rightarrow\gamma\gamma}&=&-\displaystyle{\frac{1}{3}\frac{e^2}{2}\frac{g_S}{g^2}\epsilon_i(\gamma_1)\epsilon_i(\gamma_2)}~~.
\label{eq:16}
\end{eqnarray}
We would like to make here some consideration concerning gauge invariance of the model.
This problem was dealt with in detail in \cite{hideko} in the radiative decay of axial vector mesons to a pseudoscalar and a photon.
In that case the low lying axial vectors were obtained dynamically from the interaction of a pseudoscalar and a vector within the same hidden gauge formalism used here.
Gauge invariance of the model was proved there by showing first how it works at tree level and then in the case of loops.
We follow here the same strategy.

\begin{figure}[htpb]
\begin{center}
\includegraphics[width=8cm]{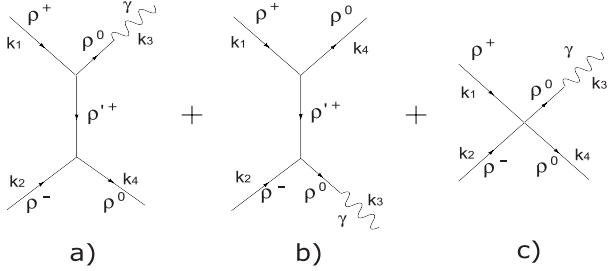}
\caption{\label{fig_new}
Feynman diagrams leading to a gauge invariance set in $\rho^+\rho^-\to\rho^0\gamma$.}
\end{center}
\end{figure}

First we show the gauge invariance of the tree level set of diagrams of fig.~\ref{fig_new} for the case $\rho^+\rho^-\to\rho^0\gamma$ (it is sufficient to make the test for one photon since for two photons it follows a fortiori).
The $\rho^0\to\gamma$ conversion proceeds via the term of eq.~(\ref{eq:14}) and, up to a constant, replaces $\epsilon_\mu(\rho^0)$ by $\epsilon_\mu(\gamma)$.
The test of gauge invariance proceeds finding a cancellation of terms upon the substitution of $\epsilon_\mu(\gamma)$ by $k_{3\mu}$, the photon momentum.
We thus proceed by substituting $\epsilon_\mu(\rho^0)$ by $k_{3\mu}$ in the strong amplitude, $\rho^+\rho^-\to\rho^0\rho^0$.
For the case of diagram a) of fig.~\ref{fig_new} we find
\begin{eqnarray}
-it^{(a)}&\equiv& i\mathcal{L}^{(a)}\nonumber\\
&=&2g^2\displaystyle{\frac{i}{k_1^2-M_\rho^2-2k_1\cdot k_3}}\{2ik_1\cdot k_3\epsilon^{+\mu}\epsilon_\mu'^+\nonumber\\
&&+ik_3\cdot \epsilon^+(k_3-k_1)^\mu\epsilon'^{+}_\mu -ik_1\cdot\epsilon^+k_3^\mu\epsilon'^+_\mu\}\nonumber\\
&\times&\{(-i\epsilon'^+_{\mu'}k_2^{\nu'}\epsilon^{-\mu'}+\epsilon'^{+}_{\mu'}(k_4-k_2)^{\nu'}\epsilon^{-\mu'})\epsilon^0_{\nu'}\nonumber\\
&&+(i\epsilon^{0\mu'}(k_2-k_4)^{\nu'}\epsilon'^{+}_{\mu'}-i\epsilon^{0\mu'}k_4^{\nu'}\epsilon'^{+}_{\mu'})\epsilon^-_{\nu'}\nonumber\\
&&+(i\epsilon^-_{\nu'}k_4^{\mu'}\epsilon^{0\nu'}+i\epsilon^-_{\nu'}k_2^{\mu'}\epsilon^{0\nu'})\epsilon'^{+}_{\mu'}\}~~.
\label{eq:16.1}
\end{eqnarray}
Upon replacing the sum of polarizations in $\epsilon'^{+}_\mu\epsilon'^{+}_{\mu'}$ leading to 
\begin{equation}
\label{eq:16.2}
-g_{\mu\mu'}+\displaystyle{\frac{(k_1-k_3)_\mu(k_1-k_3)_{\mu'}}{M_\rho^2}}
\end{equation}
we find that the second term in eq.~(\ref{eq:16.2}) leads to a vanishing contribution of eq.~(\ref{eq:16.1}).
The contribution of the diagram b) can be obtained from the one of diagram a) upon exchange $k_1\leftrightarrow k_2$, $\epsilon^+ \leftrightarrow\epsilon^-$.
The sum of polarizations for the intermediate $\rho$ meson leads to eq.~(\ref{eq:16.2}) with $k_1\to k_2$ and the contribution of the second term of the propagator vanishes equally.
Thus, only the $-g_{\mu\mu'}$ part of the propagator contributes and leads to 
\begin{eqnarray}
&&-i(t^{(a)}+t^{(b)})\nonumber\\
&=&2g^2\displaystyle{\frac{i}{2k_1\cdot k_3}}2ik_1\cdot k_3\{-2ik_2\cdot \epsilon^0\epsilon^+\cdot\epsilon^--2ik_4\cdot\epsilon^-\epsilon^+\cdot\epsilon^0\nonumber\\
&&+i(k_4+k_2)\cdot\epsilon^+\epsilon^-\cdot\epsilon^0+idem(k_1\leftrightarrow k_2, \epsilon^+\leftrightarrow \epsilon^-)\}\nonumber\\
&=&i2g^2\{2k_3\cdot\epsilon^0\epsilon^+\cdot\epsilon^--k_3\cdot\epsilon^+\epsilon^-\cdot\epsilon^0-k_3\cdot\epsilon^-\epsilon^0\cdot\epsilon^+\}\nonumber\\
\label{eq:16.3}
\end{eqnarray}
This last term provides a contribution equal, but with opposite sign, to the one of diagram c), the contact term which comes from the Lagrangian \cite{raquel} for $\rho^+\rho^-\to \rho^0\rho^0$
\begin{equation}
\mathcal{L}(\rho^+\rho^-\to\rho^0\rho^0)=2g^2(\rho^0_\mu\rho^0_\nu\rho^{+\mu}\rho^{-\nu}-\rho^0_\mu\rho^{0\mu}\rho^+_\nu\rho^{-\nu})
\label{eq:16.4}
\end{equation}
upon substitution of one $\rho^0_\mu$ by $k_{3\mu}$.
This shows that the set of diagrams of fig.~\ref{fig_new} fulfills the gauge invariance requirement.

\begin{figure}[htpb]
\begin{center}
\includegraphics[width=3.5cm]{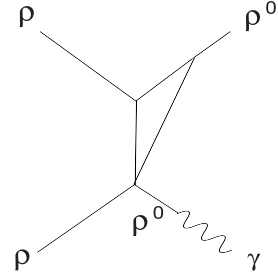}
\caption{\label{fig_newnew}
Terms encountered in the gauge invariant set of diagrams for $\rho^+\rho^-\to\rho^0\gamma$ in the case of loops.}
\end{center}
\end{figure}

The test of gauge invariance for the case of the loops contained in the dynamically generated states proceeds like in the case of the axial vector mesons by separating the intermediate propagator into its on shell and off shell parts
\begin{equation}
\frac{1}{k_1^2-M_\rho^2-2k_1k_3}=-\frac{1}{2k_1k_3}+\frac{1}{2k_1k_3}\frac{k_1^2-M_\rho^2}{(k_1-k_3)^2-M_\rho^2}
\label{eq:16.5}
\end{equation}
which allows one to take into account the on shell cancellation found before.
The rest of the terms vanish on shell and can be made to cancel a propagator.
The cancellation of terms requires now some new diagrams like the one of fig.~\ref{fig_newnew}.

Yet, the interesting thing to observe is that in all terms needed, the photon always comes from a $\rho^0$, the peculiar feature of vector meson dominance inherent in the hidden gauge formalism.
It means that the $\rho^+\rho^-\to\rho^0\rho^0$ interaction contains all these terms removing the $\gamma$ coupling.
Terms like those in fig.~\ref{fig_newnew}, with two $\rho$ mesons propagating necessarily off shell in the loops appear in the renormalization procedure of \cite{raquel} and are effectively incorporated into the scheme through renormalized couplings and subtraction constants.
As a consequence of this, the procedure followed here, coupling a $\gamma$ to any final $\rho^0$ in the strong amplitude is the right thing to do, consistent with gauge invariance.

We work in the Coulomb gauge for the photons ($\epsilon^0=0,~{\vec \epsilon}\cdot{\vec k}=0$) and to sum over the final (transverse) polarizations we use
\begin{equation}
\label{eq:17}
\sum_\lambda \epsilon_i(\gamma)\epsilon_j(\gamma)=\delta_{ij}-\frac{k_ik_j}{\vec k^2}~~.
\end{equation}
The final partial decay width, summing over final and averaging over initial state polarizations, are given by 
\begin{eqnarray}
S=2&&\nonumber\\
&&\Gamma=\displaystyle{\frac{1}{5}\frac{1}{16\pi}\frac{1}{M_R}g^2_T\frac{7}{3}\frac{1}{12}e^4(\frac{2f}{M_\rho})^4} 
\label{eq:18}\\
S=0&&\nonumber\\
&&\Gamma=\displaystyle{\frac{1}{16\pi}\frac{1}{M_R}g^2_S\frac{2}{3}\frac{1}{12}e^4(\frac{2f}{M_\rho})^4}~~.
\label{eq:19}
\end{eqnarray}
The numerical values require just the knowledge of the couplings $g_S$ and $g_T$.
Using eqs.~(\ref{eq:7}) and (\ref{eq:8}) and the results of fig.~\ref{fig:3} (where the spin projectors are excluded in $|T|^2$) we find
\begin{equation}
g^2_{S,T}=M_R\Gamma_R(|T|^2_{\rm max})^{1/2}
\label{eq:20}
\end{equation}
In tables~\ref{tab:1}, \ref{tab:2} we show the results of the couplings for different values of the $\Lambda$ used in the form factors.
\begin{table}[htbp]	
\begin{center}
\begin{tabular}{c|c|c|c|c}
$S=2$&$\Lambda$~[MeV]&$|T|^2_{\rm max}$&$\Gamma_R$~[MeV]&$g_T^2$~[MeV$^2$]\\
\hline
~~&1200&2.4$\times 10^6$&76&150$\times 10^6$\\
$M_R=1275$~[MeV]&1300&1.5$\times 10^6$&99&155$\times 10^6$\\
~~&1400&1.1$\times 10^6$&110&147$\times 10^6$\\
\end{tabular}
\caption{
Resonance parameters and coupling constants obtained by fitting the results shown in fig.~\ref{fig:3} for $S=2$ state with $q_{\rm max}=875$ MeV.
\label{tab:1}
}
\end{center}
\end{table}

\begin{table}[htbp]	
\begin{center}
\begin{tabular}{c|c|c|c|c}
$S=0$&$\Lambda$~[MeV]&$|T|^2_{\rm max}$&$\Gamma_R$~[MeV]&$g_S^2$~[MeV$^2$]\\
\hline
~~&1200&1.0$\times 10^5$&152&73.8$\times 10^6$\\
$M_R=1535$~[MeV]&1300&6.0$\times 10^4$&222&83.5$\times 10^6$\\
~~&1400&4.2$\times 10^4$&245&77.1$\times 10^6$\\
\end{tabular}
\caption{
Resonance parameters and coupling constants obtained by fitting the results shown in fig.~\ref{fig:3} for $S=0$ state with $q_{\rm max}=875$ MeV.
\label{tab:2}
}
\end{center}
\end{table}
What we can see is that, independent of the value of $\Lambda$, and hence the total width, the value of the couplings $g_S^2,~g_T^2$ is rather stable with the results
\begin{equation}
\label{eq:21}
g_S^2=78\times 10^6~~{\rm MeV}^2,~~g_T^2=150\times 10^6~~{\rm MeV}^2
\end{equation}
with uncertainties of the order of 10$\%$.
With these values, the numerical results for the $\gamma\gamma$ radiative widths are
\begin{eqnarray}
&&\Gamma(f_0(1370)\rightarrow\gamma\gamma)=1.62~~{\rm keV}\nonumber\\
&&\Gamma(f_2(1270)\rightarrow\gamma\gamma)=2.6~~{\rm keV}
\label{eq:22}
\end{eqnarray}
with estimated errors of 10$\%$.
The results for the $f_2(1270)$ are in perfect agreement with the experimental data quoted in the Introduction.
In order to compare the results obtained for the $f_0(1370)$ with experiment we need also the branching ratio $B(\pi^0 \pi^0)$ provided by the theory for this resonance. This number can be obtained from~\cite{raquel} since the total width of the $f_0(1370)$ comes about $1/4$ from $\rho \rho$ and $3/4$ from $\pi \pi$, out of which $1/4$ corresponds to $\pi^0 \pi^0$ decay.
Hence, we should compare our results of $\Gamma _{\gamma \gamma} B(\pi^0 \pi^0)= 405 $ eV with those of the Crystal Ball collaboration~\cite{Marsiske:1990hx} of  $(430\pm 80)$ eV.
The agreement is very good, but one is left to think why more accurate results are claimed in the Crystal Ball work than in \cite{uehara} in spite of having much less statistics.

The estimated 10$\%$ quoted errors are from the uncertainties in the model
parameters. This certainly does not account for the systematic uncertainties
related to how accurately the model can be a substitute for the underlying QCD
dynamics of the problem. This is obviously difficult to quantize, like  
in other
hadronic models, but should be kept in mind. Admitting that the QCD  
dynamics is
richer than the one provided by the hidden gauge mechanism used in the present
approach, the hopes are that the model resulting from the present  
framework can
be a good approximation to the real dynamics of the interaction of vector
mesons in a certain energy regime where we move. How good this  
approximation is
can only be found by testing the model with experimental data. The study done
in this work on the radiative decay has passed this test. Other tests would be
most welcome to gradually find support for the idea of these two resonances as
being, largely, dynamically generated states from the $\rho\rho$ interaction, or $\rho\rho$ bound states in the present case.
Certainly, precise measurement of the decay rate for the $f_0(1370)$ state, together with simultaneous results for both resonances in other models would be most helpful to further advance in our knowledge of the nature of these resonances.

\section{Conclusions}

We have followed recent developments in which the  $f_2(1270)$ and $f_0(1370)$ 
resonances appear as dynamically generated from the interaction of $\rho$ mesons
using the hidden gauge formalism for vector mesons. We extended the formalism
to account for the radiative decay of the resonances into $\gamma \gamma$. The
extension has been done following the standard method to deal with dynamically
generated resonances, in which the photons are coupled to the components of the
resonance, in this case $\rho \rho$.
This is technically implemented by means
of loop functions which involve the photon couplings to the components of the
resonance. In the present case, the peculiarity of the hidden gauge approach,
in which the photons couple directly to one $\rho^0$, allows a factorization of
the strong part of the interaction and the final result is converted into a
tree level contribution, hence rid of any ambiguity due to possible divergences
of the loops. The results obtained for the radiative width of the $f_2(1270)$
are in perfect agreement with experimental data.
So are those for the $f_0(1370)$ when they are compared with the experimental results of the Crystal Ball collaboration, or those of the more recent experiment by Belle within its large errors.
Yet, the large systematic errors quoted in the work from Belle, that has much better statistics, should raise some caution on these experimental numbers.
With the ultimate goal of learning about the nature of the two resonances
discussed, and having in mind the picture as dynamically generated states
emerging from the $\rho \rho$ interaction in the local hidden gauge approach, the
test passed here in the radiative decay is a first step in the search  
of support for this idea, and further tests should be most welcome.
To further strengthen this idea it would be most useful to have good
results for the radiative decay width of the $f_0(1370)$ state, as well as
results from other theoretical models for both resonances which could tell us
how stringent is the test of this radiative decay to discriminate among
different models. The work presented here should stimulate research along these
lines.

\section*{Acknowledgments}  

This work is partly supported by DGICYT contract number
FIS2006-03438. 
This research is  part of
the EU Integrated Infrastructure Initiative Hadron Physics Project
under  contract number RII3-CT-2004-506078.
The work of H. N. is supported by Japan Society for the Promotion of Science (No. 18-8661), and that of J. Y. is supported by Japan Society for the Promotion of Science (No. 19-2831).
The work of S. H. is partially supported by the Grant for Scientific Research (No. C-20540273) from Japan Society for the Promotion of Science.

\end{document}